\documentclass[12pt,amsfonts]{article}
\begin{document}
\def\p {{\partial}}
\def\n {{\nu}}
\def\m {{\mu}}
\def\a {{\alpha}}
\def\bt {{\beta}}
\def\f {{\phi}}
\def\th {{\theta}}
\def\g {{\gamma}}
\def\eps {{\epsilon}}
\def\e {{\psi}}
\def\la {{\lambda}}
\def\na {{\nabla}}
\def\k {\chi}
\def\bn {\begin{eqnarray}}
\def\en {\end{eqnarray}}
\title{Completely and Partially Integrable Systems of Total Differential Equations} \maketitle
\begin{center}
\author{Sami I. Muslih\footnote{E-mail: smuslih@ictp.trieste.it}\\International Center for Theoretical Physics(ICTP),\\Trieste, Italy}

\end{center}
\hskip 5 cm
\begin{abstract}
 Constrained Hamiltonian systems are investigated by using the Hamilton-Jacobi method. Integration of a set of
 equations of motion and the action function is discussed. It is shown that we have two types of integrable systems:
 a) ${\it Partially ~integrable ~systems}$, where the set of  equations of motion is only integrable. b) {\it Completely
 ~integrable~ systems}, where the set of equations of motion and the action function is integrable. Two examples are studied.
  \vspace{0.5cm}

\end{abstract}

\newpage

\section{Introduction}
Recently the the Hamilton-Jacobi method was initiated  [1-5] to
investigate singular system. The equivalent Lagrangian method [6]
is used to obtain the equations of motion as total differential
equations in many variables, which require the investigation of
integrability conditions.  If the system is integrable, one can
solve the equations of motion without using any gauge fixing
conditions [3,4]. In order to obtain the path integral
quantization of constrained systems [7-11], we have to discuss the
integrabilty conditions in terms of the action. In previous works
[1-5] the integrabilty conditions on the set of equations of
motion is discussed without considering the integrabilty
conditions of the action function. Also in reference [7], we
discuss the integrabilty conditions of a set of total differential
equations and the action function as well.

The aim of this paper is to show that it is not possible to obtain
the quantization for partially integrable systems and one should
consider the integrability conditions for the whole set of
equations of motion and the action function.

The plan of our paper is the following:

In section $2$, we present the Hamilton-Jacobi method. In section
$3$, integration of constrained systems is discussed. In section
$4$ two examples ar worked out and finally in section $5$ we
present our conclusions.

\section{The Hamilton-Jacobi method}

In the Hamilton-Jacobi method if we start with a singular
Lagrangian $L =L(q_i, {\dot q_{i}}, t),~i=1, 2, ..., n$, with
Hessian matrix of rank $(n-r)$, $r< n$, then the generalized
momenta can be written as

 \bn&& p_{a}= \frac{\p L}{\p
\dot q_{a}},\;\;a=1, 2, ..., n-r,\\
&& p_{\m}= \frac{\p L}{\p \dot t_{\m}},\;\;\m = n-r + 1,..., n,
\en where $q_{i}$ are divided into two sets, $q_{a}$ and $t_{\m}$.
Since the rank of the Hessian matrix is $(n- r)$, one can solve
the expressible velocities from (1) and after substituting in (2),
we get
\begin{equation}
p_{\m}= -H_{\m}(q_{i}, {\dot t_{\m}}, p_{a}; t).
\end{equation}
The canonical Hamiltonian $H_{0}$ reads
\begin{equation}
H_{0}= p_{a} {\dot q_{a}} + p_{\m} \dot{t_{\m}}|_{p_{\n}=-H_{\n}}-
L(t, q_i, \dot{t_{\n}}, \dot{q_{a}}),\;\;\;\m,\;\n=n-r+1,...,n.
\end{equation}
The set of Hamilton-Jacobi partial differential equations [HJPDE]
is expressed as [1, 2]
\begin{equation}
H^{'}_{\a}\left(t_{\bt}, q_a, \frac{\p S}{\p q_a},\frac{\p S}{\p
t_{\a}}\right) =0,\;\;\;\a,\;\bt=0, n-r+1,..., n,
\end{equation}
where $p_{\bt}= {\p S}[q_{a};t_{\a}]/{\p t_{\bt}}$ and $p_{a}= {\p
S}[q_{a};t_{\a}]/{\p q_{a}}$ with $t_{0} = t$ and $S$ being the
action.

The equations of motion are obtained as total differential
equations in many variables as follows [1,2]:

\bn
 &&dq_a=\frac{\p H^{'}_{\a}}{\p p_a}dt_{\a},\;
 dp_a= -\frac{\p H^{'}_{\a}}{\p q_a}dt_{\a},\;
dp_{\bt}= -\frac{\p H^{'}_{\a}}{\p t_{\bt}}dt_{\a}.\\
&& dz=\left(-H_{\a}+ p_a \frac{\p
H^{'}_{\a}}{\p p_a}\right)dt_{\a};\\
&&\a, \bt=0,n-r+1,...,n, a=1,...,n-r\nonumber \en where
$z=S(t_{\a};q_a)$.

\section{Completely and Partially Integrable Systems }

As was clarified, that the equations (6,7) are obtained as total
differential equations in many variables, which require the
investigation of integrabilty conditions. To achieve this goal we
define the linear operator $X_{\a}$ which corresponds to total
differential equations (6,7) as \bn X_{\a} f(t_{\bt}, q_{a},
p_{a}, z) &&= \frac{\p f}{\p t_{\a}} + \frac{\p H^{'}_{\a}}{\p
p_a}\frac{\p f}{\p q_a}- \frac{\p H^{'}_{\a}}{\p q_a}\frac{\p
f}{\p p_a} \nonumber\\&&+ (-H_{\a}+ p_a \frac{\p
H^{'}_{\a}}{\p p_a})\frac{\p f}{\p z},\nonumber\\
&&= [H^{'}_{\a}, f] - \frac{\p f}{\p z} H^{'}_{\a},\\
&&\a, \bt=0,n-r+1,...,n, a=1,...,n-r,\nonumber \en where the
commutator $[ , ]$ is the square bracket (for details, see the
appendix).

$\it{\bf lemma}$. A system of total differential equations (6,7)
is integrable if and only if
\begin{equation}
[H^{'}_{\a}, H^{'}_{\bt}] =0,\;\;\; \forall\; \a, \;\bt.
\end{equation}
$\it {\bf Proof}$. Suppose that (9) is satisfied, then \bn
(X_{\a}, X_{\bt})f &&= (X_{\a}X_{\bt} - X_{\bt}X_{\a})f,\nonumber\\
&& = [H^{'}_{\a}, [H^{'}_{\bt}, f]]- [H^{'}_{\bt}, [H^{'}_{\a},f]]
- 2 \frac{\p f}{\p z}[H^{'}_{\a}, H^{'}_{\bt}]. \en Now we apply
the Jacobi relation
\begin{equation}
[f, [g, h]] = [g, [h, f]] + [h, [f, g]],
\end{equation}
to right side of formula (10), we find
\begin{equation}
(X_{\a}, X_{\bt})f = \mathbf{[}[H^{'}_{\a}, H^{'}_{\bt}],
f\mathbf{]} - \frac{\p f}{\p z}[H^{'}_{\a}, H^{'}_{\bt}].
\end{equation}
From (9), we conclude that
\begin{equation}
(X_{\a}, X_{\bt})f =0.
\end{equation}
Conversely, if the system is Jacobi (integrable), then (13) is
satisfied for any $\a$ and $\bt$ and we get
\begin{equation}
[H^{'}_{\a}, H^{'}_{\bt}]= 0.
\end{equation}

Now the total differential for any function $F(t_{\bt}, q_{a},
p_{a})$ can be written as \bn d F && = \frac{\p F}{\p q_{a}}
dq_{a}+ \frac{\p F}{\p p_{a}} dp_{a} + \frac{\p F}{\p t_{\a}}
dt_{\a},\nonumber\\
&&= ( \frac{\p F}{\p q_a}\frac{\p H^{'}_{\a}}{\p p_{a}}-
\frac{\p F}{\p p_{a}}\frac {\p H^{'}_{\a}}{\p q_a} + \frac{\p F}{\p
t_{\a}})dt_{\a},\nonumber\\
&& =\{F, H^{'}_{\a}\} dt_{\a},\en where the commutator $\{ , \}$
is the Poisson bracket. Now, using this result, we have
\begin{equation}
dH^{'}_{\bt}= \{ H^{'}_{\bt}, H^{'}_{\a}\} dt_{\a},
\end{equation}
and, consequently, the integrabilty condition (14) reduces to
\begin{equation}
dH^{'}_{\a}=0,\;\;\forall \;\a.
\end{equation}
This is the necessary and sufficient condition that the system
(6,7) of total differential equations be completely integrable and
we call this system as ${\it Completely~ Integrable~ Model}$ .
However, equations (6) form here by themselves a completely
integrable system of total differential equations. If these are
integrated, then only simple quadrature has to be carried out in
order to obtain the action [1,2].

On the other hand, we must emphasis that the total differential
equations can be very well be completely integrable without (14)
holding and therefore without the total system (6) and (7) being
integrable and we call this system as ${\it Partially~ Integrable~
Model}$. In fact, if $\{ H^{'}_{\bt}, H^{'}_{\a}\} = F_{m}(t,
 t_{\mu})$, where $F_{m}$ are functions of $t_{\a}$ and $m$ is integer,
 then the total differential equations (6), will be integrable.

\section{Examples}
To clarify the ideas given in the previous sections, we shall
consider two examples: The first one is a partially integrable
system and the second one is a completely integrable system
\subsection{ A partially integrable system}

Let us consider, the following singular Lagrangian
\begin{equation}
L = \frac{1}{2}\left({{\dot q_{1}}^{2}} + \frac
{1}{{q_{2}}^{2}}\right),
\end{equation}
where $q_{2}\neq0$. The generalized momenta are calculated as
\begin{equation}
p_{1} = {\dot q_{1}},\;\;\;p_{2}=0.
\end{equation}
Following the Hamilton-Jacobi method, we obtain the set of
Hamilton-Jacobi partial differential equations as follows,
\begin{equation}
{H'}_{0}= p_{0} + \frac{1}{2}\left({p_{1}}^{2} - \frac
{1}{{q_{2}}^{2}}\right)=0,\;\;\;{H'}_{0}=p_{2}=0.
\end{equation}
This set leads us to obtain the following total differential
equations \bn &&dq_{1}= p_{1}dt,\;\;\;dp_{1}
=0,\;\;dp_{2}=\frac {1}{q_{2}}dt,\\
&&dz = \frac{1}{2}\left({p_{1}}^{2} + \frac
{1}{{q_{2}}^{2}}\right)dt. \en

Now, equations of motion (21) are integrable and have the
following
solutions, \bn&& q_{1} =c_{1}t + c_{2},\;\;\;p_{1}=c_{2},\\
&&p_{2}=0,\;\;\;\;\;q_{2}= c_{3}\neq0,\en where $c_{1}$, $c_{2}$
and $c_{3}$ are arbitrary constants.

Let us investigate the integrability conditions in terms of the
action. From equation (14), we obtain
\begin{equation}
\{{H'}_{2}, {H'}_{0}\}= \frac {1}{q_{2}}\neq0,
\end{equation}
One should notice that, the integrabilty conditions (25) are not
satisfied. Hence, the action function is not integrable, and it
has no unique solution.
\subsection{A completely integrable system}
As a second example, let us consider a two dimensional particle in
a uniform circular motion, the Lagrangian of this system is given
by [12]
\begin{equation}
L = \frac{1}{2} m\omega (q^{a} \epsilon_{ab}{\dot q}^{b} - \omega
q^{a} g_{ab}q^{b}).
\end{equation}
Here, $m$ is a mass parameter, $a,~b= 1, 2$, $g_{ab}$ is the
metric tensor of a two dimensional Euclidean space and
$\epsilon_{ab}$ is the completely antisymmetric tensor
$(\epsilon_{12}= + 1)$.

The canonical momenta $p_{a}$ conjugated to the generalized
coordinate $q^{a}$ are obtained as
\begin{equation}
p_{a} =- \frac{1}{2} m\omega\epsilon_{ab} q^{b}.
\end{equation}
The canonical Hamiltonian reads
\begin{equation}
H^{0}= \frac{m\omega^{2}}{2} q^{a}q_{a}.
\end{equation}
Following the Hamilton-Jacobi method, we obtain the set of
Hamilton-Jacobi partial differential equations as follows: \bn&&
{H'}^{0}= p_{0} + \frac{m\omega^{2}}{2} q^{a}q_{a}=0,\\
&&{H'}^{a}= p_{a} + \frac{1}{2} m\omega\epsilon_{ab} q^{b}=0. \en
The equations of motion and the action function are obtained as
total differential equations as follows: \bn&& dp_{a}= -
(m\omega^{2}q^{a}) dt + (\frac{m\omega}{2}\epsilon_{ab})
dq^{b},\\
&&dS =  -(\frac{m\omega^{2}}{2} q^{a}q_{a})dt  + p_{a} dq^{a}. \en
The next step is to check the integrability conditions. The total
variation of the constraint, lead us to obtain $dq^{a}$ in terms
of $dt$ as follows
\begin{equation}
dq^{a}= - (\omega\epsilon^{ab}q_{b})dt.
\end{equation}
Using (30) and (32, 33), we found after some calculations that the
integrable action has the form
\begin{equation}
S = \int dt\left[\frac{m}{2}({\dot q}_{1})^{2} -
\frac{m\omega^{2}}{2} (q_{1})^{2}\right].
\end{equation}
This result coincide with the results obtained in reference [12],
by using the Senjanovic [13] and the Batalin, Fradkin, Tyutin
$(BFT)$ [14] methods.

In fact the action in (34) is the integrable action function for
the reduced system in one dimensional harmonic oscillator.

\section{conclusion}
In this work we have investigated the integrabilty conditions in
terms of the action. In order to have a completely integrable
system (6,7), the integrabilty conditions (14) must be satisfied.
In the first  example, even though the equation of motion (21) are
integrable, the action function is not integrable and hence, it
has no unique solution. In this case we could not obtain the path
integral quatization for the model (18). For the system (26), the
integrabilty conditions lead us to obtain the integrable action in
terms of the canonical variables and this result coincide with the
results obtained in reference [12]. The obtained reduced system is
an one dimensional harmonic oscillator.

In order to obtain the path integral formulation of constrained
systems, we propose to use the action and the integrabilty
conditions on it, without any need to solve the equations of
motion. In other words, we must have a completely integrable model
in order to quantize the constrained system.

Completely integrable models are of particular interest from the
quantum point of view and hopefully, to discuss the quantization
subtleties and non- perturbative effects [15-17]. These results
encourage us to investigate the integrable models in
two-dimensional $Dilaton~ Gravity$ by using the $Hamilton-Jacobi$
method, and this study is now under investigation.
\appendix

\section{Square brackets and Poisson brackets}
In this appendix we shall give a brief review on two kinds of
commutators: the square and the Poisson brackets.

The square bracket is defined as
\begin{equation}
[F, G]_{q_{i}, p_{i}, z}= \frac{\p F}{\p p_{i}} \frac{\p G}{\p
q_{i}}- \frac{\p G}{\p p_{i}}\frac{\p F}{\p q_{i}} + \frac{\p
F}{\p p_{i}}(p_{i} \frac{\p G}{\p z}) - \frac{\p G}{\p
p_{i}}(p_{i} \frac{\p F}{\p z}).
\end{equation}
The Poisson bracket is defined as
\begin{equation}
{\{f, g\}}_{q_{i}, p_{i}}= \frac{\p f}{\p p_{i}} \frac{\p g}{\p
q_{i}}- \frac{\p g}{\p p_{i}}\frac{\p f}{\p q_{i}}.
\end{equation}
According to above definitions, the following relation holds
\begin{equation}
[H^{'}_{\a}, H^{'}_{\bt}]= \{H^{'}_{\a}, H^{'}_{\bt}\}.
\end{equation}
 \section*{Acknowledgments}

The author would like to thank ICTP for support and hospitality
during this work.

\end{document}